\documentclass{article}
\usepackage[utf8]{inputenc}
\usepackage{amsfonts}
\usepackage{spconf,amsmath,graphicx}
\usepackage{xcolor}
\graphicspath{{./figures/}}
\usepackage{subfigure}
\usepackage[T1]{fontenc}
\usepackage{csquotes}
\usepackage{float}
\usepackage{bm}

\usepackage{comment}
\usepackage{amsmath,amssymb,mathtools, commath}
\usepackage{multicol}
\usepackage{booktabs}
\usepackage{tabularx}
\usepackage{multirow}
\usepackage{balance}
\usepackage{pgfplots}
\usepackage[noadjust]{cite}
\usepackage[caption=false,font=normalsize,labelfont=sf,textfont=sf]{subfig}
\usepackage{tikz}

\pgfplotsset{compat=newest}

\usetikzlibrary{positioning,spy,backgrounds,calc,arrows}
\graphicspath{{./images/}}

\newcommand{\bs}{\boldsymbol}
\newcommand{\mbf}{\mathbf}

\title{Deep Image Prior using Stein's Unbiased Risk Estimator: SURE-DIP}
\name{Maneesh John$^*$, Hemant Kumar Aggarwal$^*$, Qing Zou, and  Mathews Jacob 
	\thanks{This work is supported by 1R01EB019961-01A1 and 1 R01 AG067078-01A1. This work was conducted on an MRI instrument funded by 1S10OD025025-01. $*$ Both authors contributed equally to the work. }
	}
\address{The University of Iowa}%

\begin{document}
%\ninept
%
\maketitle
\begin{abstract}

Deep learning algorithms that rely on extensive training data are revolutionizing image recovery from ill-posed measurements. Training data is scarce in many imaging applications, including ultra-high-resolution imaging. The deep image prior (DIP) algorithm was introduced for single-shot image recovery, completely eliminating the need for training data. A challenge with this scheme is the need for early stopping to minimize the overfitting of the CNN parameters to the noise in the measurements. We introduce a generalized Stein's unbiased risk estimate (GSURE) loss metric to minimize the overfitting. Our experiments show that the SURE-DIP approach minimizes the overfitting issues, thus offering significantly improved performance over classical DIP schemes. We also use the SURE-DIP approach with model-based unrolling architectures, which offers improved performance over direct inversion schemes.

\end{abstract}
\begin{keywords}
 single shot, image reconstruction, SURE, deep image prior
\end{keywords}
\section{Introduction}

The reconstruction of images from a few noisy measurements is a central problem in several modalities, including MRI, computer vision, and microscopy. Classical methods, including compressed sensing (CS), that exploit image priors have made great strides in the past. Recently, deep learning algorithms have emerged as powerful alternatives offering improved performance over CS-based methods that often rely on carefully handcrafted regularization priors. Most deep learning methods for image reconstruction rely on learning of trainable convolutional neural network (CNN) modules within the network using fully sampled training images. In addition to computational efficiency, these deep-learning-based methods provide better image quality than classical CS-based approaches. However, one of the main challenges in many application areas (e.g., high-resolution applications, microscopy of organelles that are imaged for the first time) is the scarcity of training images. 

The deep image prior (DIP) approach was introduced to eliminate the need for training data \cite{dip2018}. Here, the image is modeled as the output of a CNN generator $\boldsymbol x = f_{\Phi}(\boldsymbol u)$. Here $\boldsymbol u$ is an arbitrary input to the generator $f_{\Phi}$ and $\mathcal A$ is the linear forward model. The CNN parameters denoted by $\Phi$ are optimized such that the measurements $\mathcal A (f_{\Phi}(\boldsymbol u))$ of the image closely match the actual measurements. This approach uses the inherent bias of CNN architectures to image content, thus regularizing the recovery from few measurements. The benefit of this scheme is that it can be readily applied to imaging inverse problems, where training data is not available. A challenge with these schemes is the need for careful early stopping. Specifically, CNNs often have sufficient capacity to learn measurement noise. Without early stopping, the algorithm will over-fit the noise in the measurements, resulting in poor image quality. Regularization strategies, including adding noise to the weights during training \cite{cheng2019bayesian}, have been introduced to minimize this overfitting issue. 

The main focus of this paper is to systematically address the overfitting issue in DIP using the projected Stein's unbiased risk estimator (SURE)~\cite{sure,eldarGSURE} criterion. The SURE metric is an unbiased estimator for mean-square-error (MSE) \cite{sure} and is widely used in image denoising to select regularization parameters and to train CNN networks for image denoising \cite{zhussip2019cvpr}. Because only partial information about the image, specified by the forward model $\mathcal A$, is available, the traditional SURE approaches are not directly applicable to our setting. An alternative is \cite{metzler2018}, where the SURE approach is used in each step of an unrolled algorithm to train the image denoiser; this approach makes the approximation that the images at each unrolled stage are corrupted by Gaussian noise. We instead propose to use the projected generalized SURE (GSURE) approach \cite{eldarGSURE} to address the noise overfitting issue. In particular, we use the GSURE metric to approximate the true mean-square error in the range space of the $\mathcal A$ operator. The GSURE metric consists of a data consistency term, which involves the comparison of the measurements of the image to the true measurements, and an extra divergence of the network $f_{\Phi}$ that accounts for the noise in the measurements. The divergence term acts as a regularizer on the network, thus minimizing the risk of overfitting. The CNN parameters are initialized randomly and fitted to the measurements of the image; the proposed approach is thus similar to conventional sparse optimization schemes that do not require pre-training of the network components.  

Unlike traditional DIP schemes, we also consider unrolled model-based architectures \cite{modl}. In particular, we hypothesize that these unrolled architectures, which encourage data consistency, are more efficient in the inverse problem setting than the simpler UNET models used in the traditional DIP setting \cite{dip2018}. The image reconstruction scheme is similar to the training strategy in unrolled schemes \cite{modl}, except that the training is performed using undersampled measurements of a single image with the additional divergence term. Our experiments show that the unrolled architectures offer improved performance and faster convergence compared to the traditional UNET schemes. An ensemble-SURE approach \cite{ensure} recently presented  uses the SURE approach to train a model from a large ensemble of undersampled images, acquired using different measurement operators, with the goal of applying it to a test image without re-training. By contrast, the focus of this paper is on single-shot learning, where the model is learned from the undersampled data of a single image. We note that the GSURE metric may also be used to adapt a pre-trained model to a new setting, which is also a problem of high importance.

\section{Proposed Method}
The image acquisition model to acquire the noisy and under-sampled measurements $\bs y \in \mathbb C^n$ of an image $\bs x \in \mathbb C^m$ using the forward operator $\mathcal A$ can be represented as
\begin{equation}
	\label{eq:fwd}
	\bs y=\mathcal A (\bs x) +\bs n
\end{equation}
We assume that noise $\bs n$ is Gaussian distributed with mean zero and standard deviation $\mbf \sigma$ such that $\bs{n}\sim \bs N(0,\mbf \sigma)$. We consider the approximate reconstruction {$\bs u=\mathcal A^H (\bs y)$}, which lives in the range space of $\mathcal A$ specified by $\mathcal V$, as the input to the CNN.  The DIP recovery using a deep neural network $f_\Phi$ with trainable parameters $\Phi$ is represented as
\begin{equation}
	\label{recon}
	\bs{\widehat x} = f_{\Phi}(\bs {u}). 
\end{equation}
Here $f_\Phi$ can be a direct-inversion or a model-based deep neural network such as \cite{modl}, whose parameters are denoted by $\Phi$. If the ground-truth training images $\bs x \sim \mathcal M$ are available, one can train the parameters of the network using 
\begin{equation}
	\label{eq:mse}
	\text{MSE}= \mathbb E_{\bs x \sim \mathcal M}~ \| \bs{\widehat x} -  \bs x \|_2^2.
\end{equation}
In this work, we assume that ground truth images are not available, and the recovery is performed using the measurements \eqref{eq:fwd} of a single image. The DIP approach optimizes the parameters $\Phi$ with the loss function:
\begin{equation}
	\label{eq:dip}
	\text{DIP}(\Phi)= \| \mathcal A\left(f_{\Phi}(\bs u)\right) -  \bs y \|_2^2.
\end{equation}
Here, the CNN parameters $\Phi$ are initialized with random values and are optimized such that \eqref{eq:dip} is minimized. We note that CNN models often have high representation power; they can represent noise when trained with adequate epochs. Because the measurements $\bs y$ in \eqref{eq:fwd} are noisy, the DIP scheme is vulnerable to overfitting. Early termination is used in \cite{dip2018} to minimize the risk of overfitting. 

\begin{figure}	\centering
	\subfigure[data-term] {	
		\includegraphics[ width=.3\textwidth]{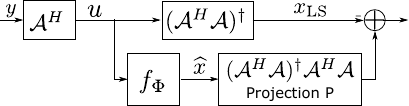}	
	}
	
	\subfigure[divergence-term] {	
		\includegraphics[width=.3\textwidth]{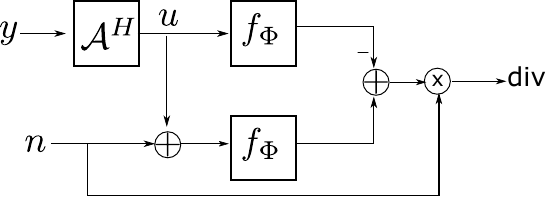}	\vspace{-2em}
	}\vspace{-1em}
	\caption{\footnotesize{The implementation details of the GSURE-based loss function for model adaptation. (a) shows the calculation of the data-term. (b) shows the calculation of the divergence term. Here we pass the regridding reconstruction and its noisy version through the network and find the error between the two terms. Then we take the inner product between this error term and the noise to get an estimate of the network divergence divergence.}}\vspace{-1em}
	\label{fig:gsure} 
\end{figure}

\begin{figure*}[t!]
\begin{center}
\subfigure[PSNR vs epochs]{\includegraphics[ height=0.3\textheight]{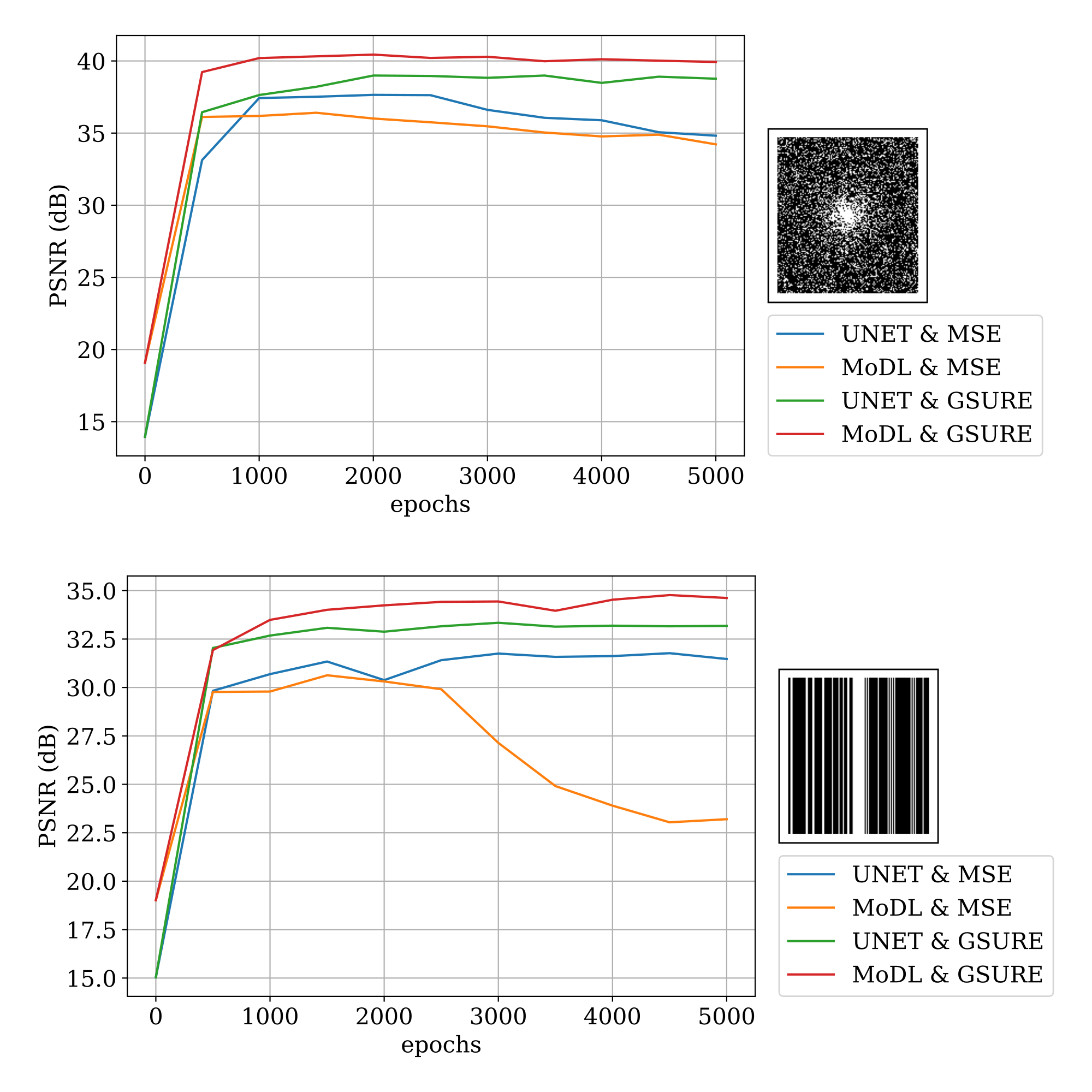}}
\subfigure[Images recovered using SURE-DIP and MSE-DIP]{\includegraphics[ height=0.3\textheight]{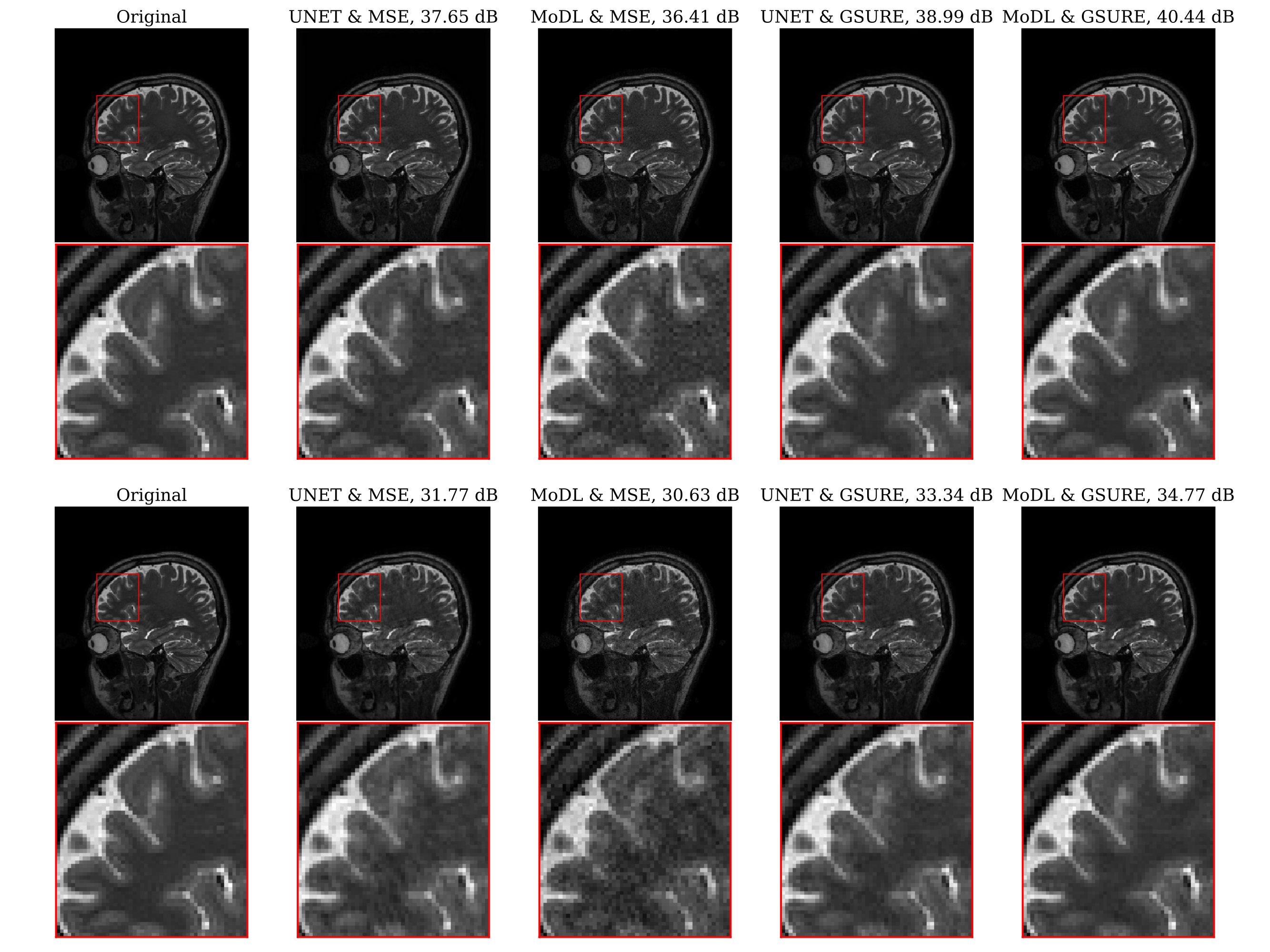}}\vspace{-1em}
	\caption{Single-shot DIP using different losses and architectures. We train the UNET and the MoDL architectures using the MSE and G-SURE losses. (a) shows the evolution of the peak signal to noise ratio (PSNR) with epochs. The top row corresponds to a 4x acceleration using a non-uniform mask, while the bottom row corresponds to the more challenging setting of 4x acceleration using a 1-D mask. (b) shows the images recovered by the methods, which offer the best PSNR, corresponding to the maxima in the plots in (a). }\vspace{-1em}
	\label{sure_vs_dip} 
\end{center}
\end{figure*}

%\begin{figure*}[t!]
%\includegraphics[ %width=0.5\textwidth]{figures/dip-sure-fig2-outs.png}	
%	\caption{\footnotesize{SURE vs DIP.}}
%	\label{sure_vs_dip_outs} \end{figure*}

We propose to use projected GSURE~\cite{eldarGSURE} loss function that explicitly accounts for the noise in the measurements to minimize overfitting issues. The GSURE loss approximates the projected MSE (PMSE)
\begin{equation} \label{pmse}
	{\rm PMSE}(\Phi) =~  = \mathbb E_{\bs n}\left [ \|\mbf P \left(f_{\Phi}(\bs u)\right) -\mbf P (\bs x)  \|_2^2 \right].
\end{equation}
Here, $\mbf P$ is the projection operator to the range space of $\mathcal A$, denoted by 
\begin{equation}
\mathbf P(\bs z) =  (\mathcal A^H \mathcal A)^{\dag}\mathcal A^H \mathcal A~\bs z \approx \lim_{\lambda\rightarrow 0}(\mathcal A^H \mathcal A+\lambda \mathcal I)^{-1}\left(\mathcal A^H \mathcal A\right)~\bs z
\end{equation}
The PMSE is only an approximation for the MSE in \eqref{eq:mse}, since it cannot measure the parts of the signal in the null space of $\mathcal A$. Unfortunately, one cannot directly measure PMSE as specified by \eqref{pmse} when $\bs x$ is not known. If the measurements in \eqref{eq:fwd} were not noisy (i.e, $\bs n=0$), one could obtain the projected image as $\mathbf P(\bs x) = \bs x_{\rm LS} = (\mathcal A^H \mathcal A)^{\dag}\mathcal A^H \bs y$. However, the measurements in \eqref{eq:fwd} are corrupted by noise, which makes it impossible to directly evaluate PMSE in \eqref{pmse}.
%\begin{equation}
    
%\end{equation}
The GSURE metric \cite{eldarGSURE} is an unbiased estimate for the projected MSE, denoted by $\|\mathbf P(\bs{\widehat x} - \bs x)\|^2$: 
\begin{equation} \label{gsure}
	\mathcal L =~ \underbrace{\mathbb E_{\bs u} \left [ \|\mbf P (\bs{\widehat x}) -\bs x_{\text{LS}}  \|_2^2 \right]} _{\mathrm{data~ term}} ~+
	 ~\underbrace{2 \mathbb E_{\bs{u}} \left [\nabla_{\bs{u}} \cdot  f_\Phi(\bs{u})  \right ]}_ {\mathrm{divergence}}. 
\end{equation}

The first term in \eqref{gsure} generalizes the loss in \eqref{eq:dip}. In particular, for simple forward operators such as inpainting, the data term in \eqref{gsure} simplifies to \eqref{eq:dip}. The second term is a measure of the divergence of the network and is computed using the Monte Carlo approach \cite{mcsure}. One may view this term as a network regularization term, which regularizes the parameters of the deep CNN, thus minimizing the risk of overfitting. The SURE-based approach can be seen as an alternative to the noise addition approach in \cite{cheng2019bayesian}, weight regularization in \cite{gstorm}, and early stopping in \cite{dip2018}.

\section{Experiments and Results}
We demonstrate the SURE settings in the context of single-shot MR image recovery. However, we note that the proposed framework is readily applicable to similar single-shot problems in image inpainting, deblurring \cite{dip2018}. We consider publicly available~\cite{modl} parallel MRI brain data. The matrix dimensions were $256\times256\times208$ with a 1~mm isotropic resolution. The dataset consists of fully sampled multi-channel brain images from nine volunteers, out of which the data from five subjects were used for training of supervised strategies, the data from two subjects were used for testing, and the data from the remaining two subjects were used for validation. The k-space data was normalized so that the real and imaginary values of the images are scaled between -1 and 1. We note that the measurement data is already noisy. However, to test the impact of noise on the proposed scheme, we add additional noise with noise standard deviation ($\sigma=0.01$) to the measurements. In the single-shot SURE-DIP setting, we optimize the parameters of both a UNET architecture and an unrolled architecture with ten unrolling steps using the publicly available implementation \cite{modl} using the SURE metric. A five-layer network, whose parameters are shared across iterations, was used in the unrolled setting.

\begin{figure*}[t!]
\includegraphics[ width=1.0\textwidth]{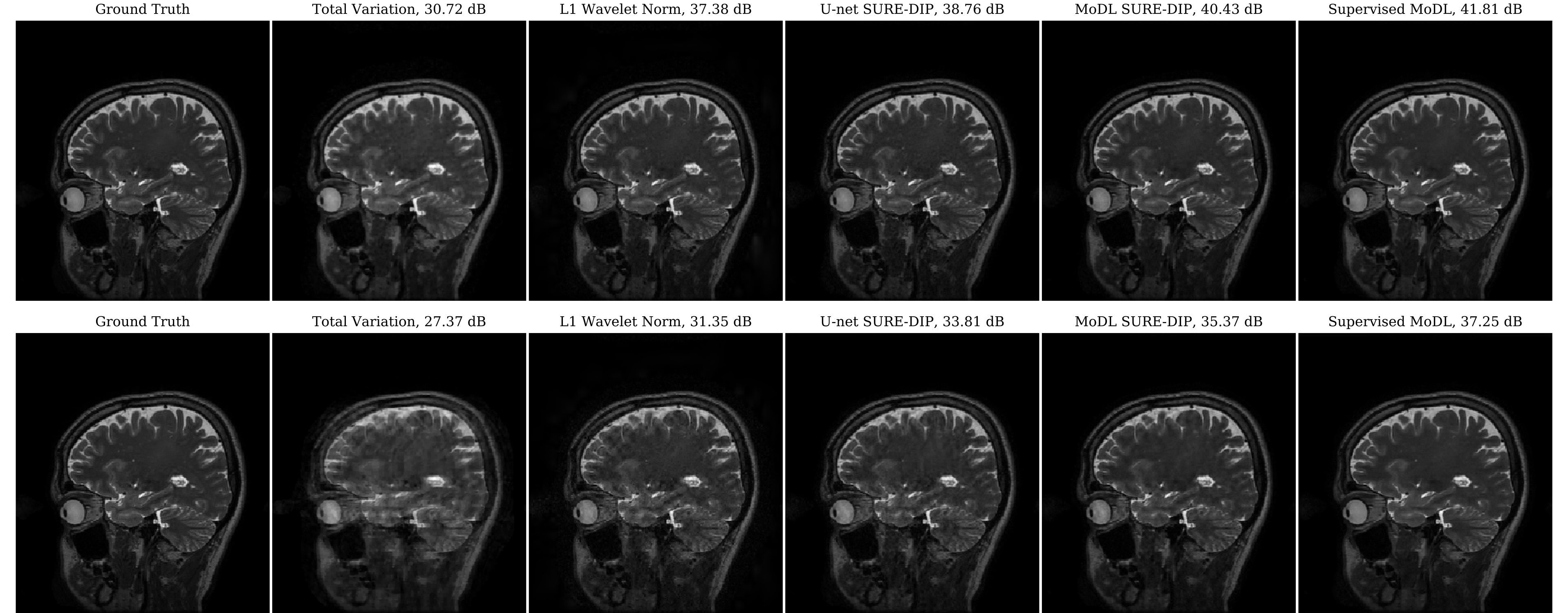}
	\caption{\footnotesize{Comparison of SURE-DIP trained networks (UNET and unrolled) against state-of-the-art single-shot methodsas well as an unrolled deep learning scheme that was trained on exemplar data. The top row corresponds to a 4x acceleration using a 2D mask, shown in Fig. 2, while the bottom row corresponds to a 4x acceleration using a 1D mask (see Fig. 2). The results show that the SURE-DIP schemes can offer improved performance over single-shot methods, while the performance is marginally lower than trained MoDL. }}\vspace{-1em}
	\label{fig:sota} 
\end{figure*}

\subsection{Comparison with DIP}
We first compare the proposed SURE-DIP metric in \eqref{gsure} against the MSE \eqref{eq:dip} metric in Fig. \ref{sure_vs_dip}. We consider both the UNET and the unrolled architectures in the 4x undersampled setting using a 2D undersampling. Our experiments show that the use of the $\ell_2$ loss metric in the measurement domain peaks and then gradually degrades as reported in \cite{dip2018}. The degradation can be viewed as overfitting of the model parameters to the noise in the measurements. By contrast, the PSNR of the models trained using SURE metric are observed to almost monotonically improve with epochs. The improvement can also be appreciated from reduction in noise amplification, visible in the reconstructed images.

\subsection{Comparison with state-of-the-art methods}
We compare the proposed scheme against state-of-the-art single-shot methods (TV regularization and $\ell_1$ Wavelet regularization) as well as a deep CNN (MoDL \cite{modl}) trained using exemplar data in Fig. \ref{fig:sota}. We perform the reconstructions for two different sampling patterns, both corresponding to 4x undersampling. The experiments show that the proposed scheme offers significantly improved results compared to the classical single-shot methods. It performs marginally worse compared than supervised-MoDL, which is expected. 

\vspace{-1em}
\section{Discussion and Conclusions}
We introduced the SURE metric for single-shot training of deep-learning imaging reconstruction algorithms, including DIP. The SURE metric systematically accounts for the Gaussian noise in the measurements, thus minimizing the risk of overfitting of the model. The experiments show that the proposed metric offers significantly improved image quality compared to the traditional MSE metric. In particular, the SURE metric eliminates the need for early stopping that is often needed in DIP schemes. The comparison against state-of-the-art single-shot schemes show that the proposed scheme offers significantly improved performance, while the performance is marginally lower than that of the trained MoDL scheme. The proposed approach is applicable to problems where training data is not available or is difficult to acquire.

\section{Compliance with Ethical Standards}
This research study was conducted using publicly available human subject data. Ethical approval was not required as confirmed by the license attached with the open access data.
\vspace{-1em}
\bibliographystyle{IEEEbib}
\bibliography{bibTexSamp}

\end{document}